\documentclass[%
reprint,
 amsmath,amssymb,
prapplied,
]{revtex4-2}

\usepackage{graphicx}
\usepackage{dcolumn}
\usepackage{bm}
\usepackage{float}
\usepackage{color}
\usepackage[caption=false]{subfig}
\usepackage{siunitx}
\usepackage{tabularx}
\usepackage{array,multirow}
\usepackage[thinlines]{easytable}

\begin{document}

\title{Harnessing the frequency eigenchannels of ultrafast time--varying media}

\author{I. R. Hooper}
\email{email: i.r.hooper@exeter.ac.uk}
\affiliation{%
Department of Physics and Astronomy, University of Exeter, Stocker Road, 
Exeter, Devon, UK, EX4 4QL\\
}%
\author{D. B. Phillips}
\affiliation{%
Department of Physics and Astronomy, University of Exeter, Stocker Road, 
Exeter, Devon, UK, EX4 4QL\\
}%
\author{S. A. R. Horsley}
\affiliation{%
Department of Physics and Astronomy, University of Exeter, Stocker Road, 
Exeter, Devon, UK, EX4 4QL\\
}

\begin{abstract}
Control over the interaction of waves with ultrafast time-varying materials -- those that change on a time-scale commensurate with the wave period -- holds much promise for developing a raft of new technologies.
Time-varying materials exchange energy with the waves passing through them, thus exhibiting entirely new phenomena not possible with static media.  Work to date has largely considered the realisation of ultrafast time-varying materials, with much less attention paid to how their interaction depends upon the shape of the incident waves themselves.  Here we experimentally demonstrate \emph{waveform shaping} within ultrafast time-varying media: temporal shaping of incident pulses to optimise specific dynamic material interactions.  We create an ultrafast time-varying medium by rapidly modulating a waveguide termination, and measure the spectrally resolved reflection matrix.  Analysis of this reflection matrix enables the identification of the incident waveforms that undergo a desired spectral transformation. Using this approach, we first characterise the medium's eigenpulses; temporal modes that reflect from the medium without spectral distortion. Next, we identify the waveforms that are either maximally or minimally reflected, enabling strong broadband absorption arising from the temporal modulation of the medium. Finally, we show how energy can be optimally concentrated into a user-defined set of frequency bands upon reflection.  Our work paves the way towards new applications of ultrafast time-varying media, spanning from optics to microwaves and acoustics.
\end{abstract}

\maketitle

Many materials we encounter exhibit time-varying electromagnetic
properties: a cloud momentarily obscuring the sun, the rippling caustics cast on the floor of a swimming pool, or the fading of radio signals as atmospheric conditions change. But in most cases, there is a vast separation of scales between such variations and the frequency of the electromagnetic wave itself.  While visible light oscillates nearly a quadrillion times per second, the characteristic motion of matter---due to acoustic or mechanical vibrations---typically occurs at kilohertz frequencies.  Recent ultrafast experiments~\cite{alam2016,zhou2020,bohn2021,tirole2022,lustig2023} have begun to bridge this frequency gap, dynamically modulating a medium's refractive index by order--unity amounts within just a few wave cycles.  This progress has inspired
a surge of experimental and theoretical investigations into the physics of wave propagation in ultrafast time--varying materials~\cite{Galiffi2022}. 

The physical significance of such rapid changes in material properties can be appreciated by comparing to the equivalent situation in space.  A sharp spatial interface between two materials induces wave scattering into different \emph{momenta}.  Yet, when the refractive index changes abruptly in time, `scattering' means the generation of new \emph{frequencies}.  For a spatially uniform temporal modulation, where momentum is conserved, additional waves are generated with the negative of the initial wave frequency, leading to a time reversal of the field~\cite{Mendonca2000,mendoncca2002,bacot2016,moussa2023}.  Meanwhile, for a rapid periodic temporal modulation of the refractive index there is the analogue of the usual spatial diffraction and frequency band structure~\cite{joannopoulos1997,BUTT2021}, where momentum band structure emerges~\cite{lustig2023,Wang2023}, with associated topological indices~\cite{ren2025}.  However, within these band gaps, rather than propagation being forbidden the field undergoes parametric amplification, leading to an exponential increase in the field amplitude. In the same spirit, recent experimental work has demonstrated the temporal equivalent of Young's double-slit experiment \cite{tirole2023} as well as coherent perfect absorption and amplification \cite{galiffi2023,galiffi2024}.

There are a large number of new results in this rapidly growing field that we cannot do justice to here. It is important to note that many have no simple spatial analogue like those described above, such as the potential to violate established bounds on device performance~\cite{Hayran2023}, temporal aiming~\cite{pacheco-pena2020}, and synthetic motion through combined space--time modulation~\cite{pendry2021,harwood2025}.  General review and tutorial style articles on time-varying and space-time varying media can be found in~\cite{Caloz2020,Caloz2020-2,Galiffi2022}.

In this work we explore another analogy between spatially and temporally inhomogeneous materials. We demonstrate the temporal analogue of spatial wavefront shaping. In the spatial domain, wavefront shaping enables precise control of the propagation of optical fields through highly complex, often disordered, spatially scattering media~\cite{gigan2022roadmap}.

\begin{figure*}[t!]
    \includegraphics[width=0.95\textwidth]{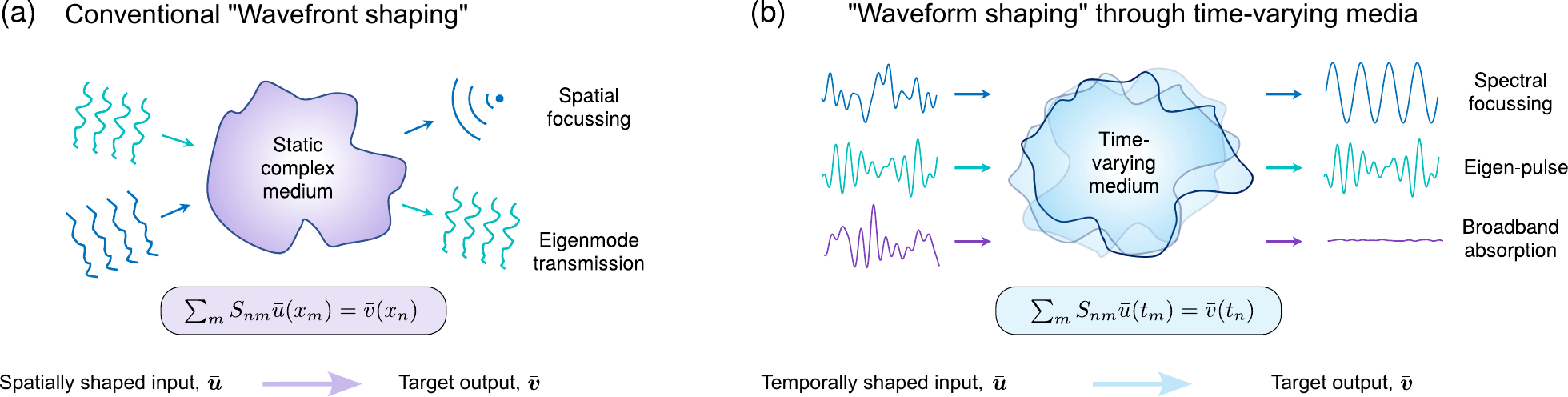}
    \caption{\textbf{From spatial `wavefront shaping' to temporal `waveform shaping'}. \textbf{(a)} Wavefront shaping through static spatially scattering complex media. Once the spatial scattering matrix of a static medium, $\boldsymbol{S}$, has been characterised, it is used to find specific input fields, $\bar{\boldsymbol{u}}$ (here given at a set of positions $x_{m}$), that will evolve into target output fields, $\bar{\boldsymbol{v}}$, after propagation through the medium. Above we show two examples of input wavefronts shaped to generate different target output fields at the far side of the medium: a focussed spot, and an eigenmode that preserves the spatial shape of the field upon transmission. \textbf{(b)} Waveform shaping through a time-varying medium. Analogous to the spatial case, once the temporal scattering matrix, $\boldsymbol{S}$, of the medium has been characterised, it can be used to find specific input waveforms, $\bar{\boldsymbol{u}}$ (here given at a set of times $t_m$), that will evolve into target output waveforms $\bar{\boldsymbol{v}}$ after propagation through the time-varying medium. Above we show three examples of input waveforms shaped to generate different target output waveforms after interaction with the same medium: a `spectrally focussed' single frequency output, an `eigenpulse' that preserves the temporal shape of the waveform upon transmission, and an input that is almost entirely absorbed.}
    \label{Fig_1}
\end{figure*}

Indeed, it is well known that spatial disorder gives rise to rich behaviour, such as Anderson localisation~\cite{Anderson1958,Chabanov2001}, and a vibrant field has emerged around techniques for imaging through such complex scattering media~\cite{Vellekoop2007,Rotter2017,bertolotti2022,mididoddi2025}.  Wave propagation in random time-varying media is less understood, though recent theoretical work has predicted a temporal analogue of Anderson localisation~\cite{Carminati2021,Eswaran2025}, and extensions of the techniques used in spatial wavefront shaping to time-varying media~\cite{Globosits2024}. To date, experimental investigations of randomly time-varying media have been limited to the case of slow temporal modulations~\cite{mididoddi2025}.

Taking inspiration from advances in spatial wavefront shaping, here we undertake the first experimental demonstration of a technique we coin 
{\it waveform shaping}, where a wave is shaped in the time domain to optimise a selected interaction with an ultrafast time-varying medium. This allows both the dissipation and dispersion to be enhanced or suppressed in a specified way.

We identify waveforms that are spectrally unchanged when propagated through our time-varying medium---so-called eigenpulses~\cite{Horsley2023}---and those waveforms that are maximally and minimally reflected, demonstrating broadband absorption as a result of the temporal modulation. Furthermore, we `focus' an incident waveform to specific narrow frequency bands with high contrast.  Unlike a conventional filter, which is a static device that absorbs or reflects a band of frequencies, transmitting only e.g., a narrow band, this frequency `focussing' is achieved through \emph{converting} the spectral content into a specified band.  More generally, our work demonstrates how the interactions between waves and ultrafast time-varying media can be optimised to maximise or minimise any desired effect.\\

\noindent{\bf From spatial wavefront shaping to temporal waveform shaping}

\noindent The sheer complexity of the multiple scattering of light was long thought to uncontrollably fragment optical fields, irreparably scrambling any spatial information they carry. Yet the technique of wavefront shaping has recently emerged as an effective, versatile way to control light within highly scattering environments~\cite{Vellekoop2007}; examples including frosted glass, multimode optical fibres or biological tissue~\cite{mosk2012controlling}.  Wavefront shaping makes use of the characterisation of materials in terms of their scattering matrix, $\boldsymbol{S}$.  This object serves as an intricate model of the complex medium's effect on light~\cite{popoff2010measuring}.  To calculate the effect of the medium, we choose a basis (e.g. the Fourier, or real space basis) in which to represent the scattering matrix.  The scattering matrix then tells us the complex weighting of each basis function, representing the outgoing field (these amplitudes written as the vector $\boldsymbol{v}$) in terms of the amplitudes of the incoming field (written as the vector $\boldsymbol{u}$).  Assuming the material response is linear, this relationship is,
\begin{equation}
    \boldsymbol{v}=\boldsymbol{S}\cdot\boldsymbol{u}.\label{eq:scattering-matrix}
\end{equation}
For instance, when a monochromatic wave is incident onto a rough transparent interface, such as a piece of frosted glass, the wave will be scattered into a range of momenta in both transmission and reflection. An appropriate basis for $\boldsymbol{u}$ and $\boldsymbol{v}$ is the Fourier basis, where each vector element represents the complex amplitude of a component wave propagating in a single direction.  The element $S_{mn}$ of the scattering matrix then describes the complex amplitude of the wave scattered into output direction $m$ upon coherent illumination with input direction $n$. 


Knowledge of either the full scattering matrix $\boldsymbol{S}$ or one of its sub--matrices reveals how input field $\boldsymbol{u}$ should be pre-shaped such that, upon interaction with even an incredibly complex scattering environment, the field evolves into a user-defined output state $\boldsymbol{v}$.  For example to generate a focus at the output, $\boldsymbol{v}=\boldsymbol{v}_{\rm Focus}$, the input $\boldsymbol{u}_{\rm Focus}=\boldsymbol{S}^{-1}\cdot\boldsymbol{v}_{\rm Focus}$ is sent into the system and, via Eq. (\ref{eq:scattering-matrix}), this will generate the desired output field (See Fig.~\ref{Fig_1}).  We can also work with the singular value decomposition (see e.g.~\cite{choi2011}), or the eigen--decomposition of the scattering matrix, using knowledge of the singular values or eigenvalues to control both the amount of energy lost in passage from input to output, and the distortion of the signal.  Such scattering matrix-based approaches have been widely applied to imaging, sensing and information transfer through opaque media~\cite{cao2022shaping}.  These methods have also been extended to account for the wavelength dependence of the scattering matrix, via measurement of the multispectral transmission or reflection matrix~\cite{mounaix2016spatiotemporal,zhang2023deep,balondrade2024multi}.


 The vast majority of wavefront shaping approaches have considered static media, the material unvarying at least over the time taken to measure and apply the scattering matrix.  Such cases always exhibit a multispectral scattering matrix without coupling between input modes of differing wavelength.  Here we shall extend these ideas to ultrafast time-varying media (TVM), where there is usually strong frequency--frequency coupling.  The concept of a wavefront is not obviously applicable in this general case and we thus dub the technique ``waveform shaping''.
 
 We consider the simplest case, replacing a medium that couples many different spatial modes with one coupling temporal modes.  By analogy with a static complex medium, which is inhomogeneous in \emph{space}, this time-varying medium is inhomogeneous in \emph{time}.  Instead of scattering the wave into a range of momenta, the temporal inhomogeneity couples different frequencies of radiation.
 
\begin{figure}[ht]
    \includegraphics[width=0.48\textwidth]{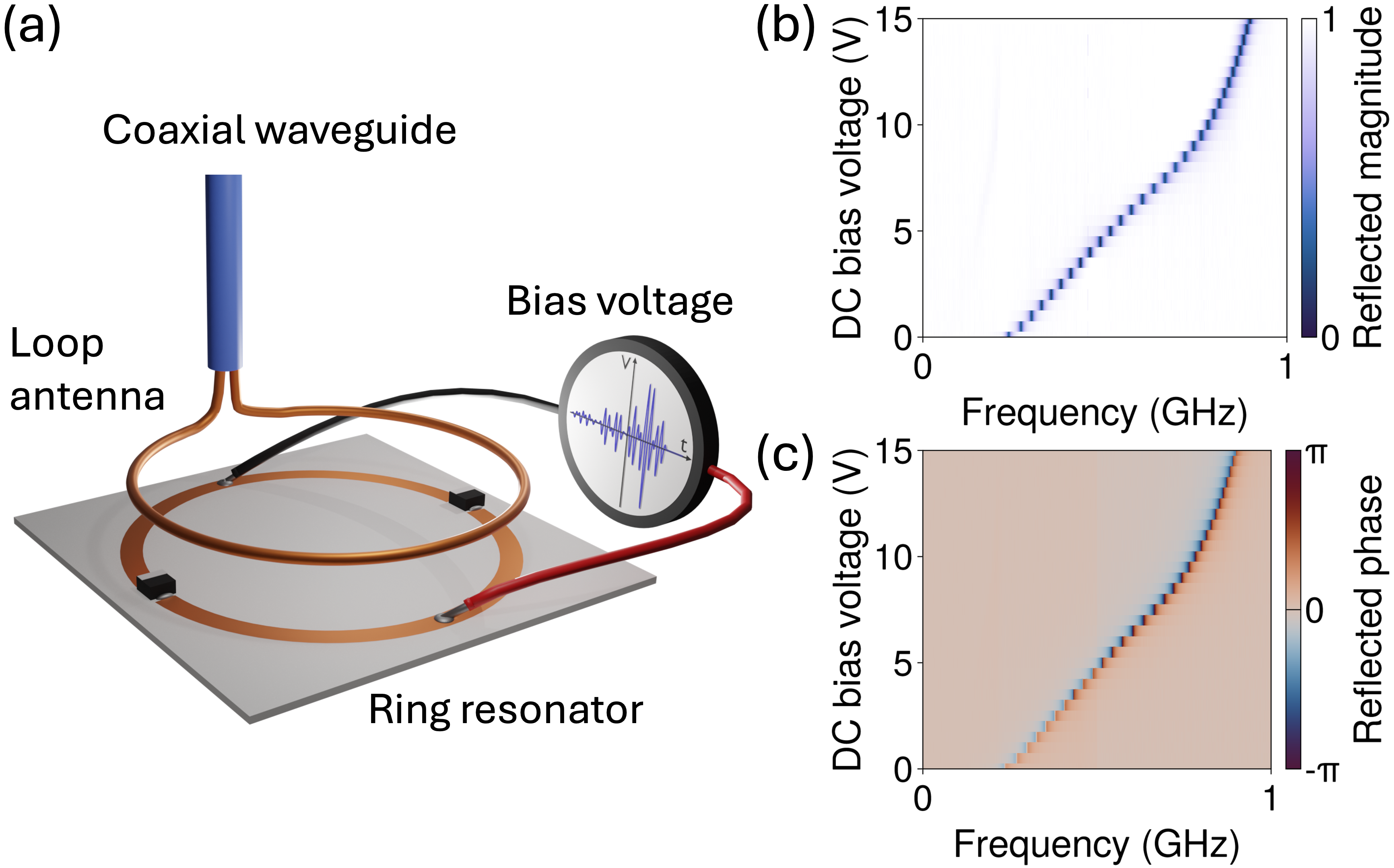}
    \caption{\textbf{A transmission line with a time-varying termination.} \textbf{(a)} A ring resonator with varactor diodes across two gaps is inductively coupled to a coaxial transmission line via a loop antenna, the coupling being controlled via the distance between them. \textbf{(b)} \& \textbf{(c)} By applying a reverse bias voltage across the varactors, their capacitance can be changed as a function of time, modifying the resonance frequency. Here we show the reflected magnitude (b) and phase (c) as a function of frequency and applied DC bias voltage for the case where the resonator is `ideally' coupled - i.e., the gap between the resonator and the loop antenna is such that the reflectance is minimal on resonance.}
    \label{Fig_2}
\end{figure}

 In accordance with the experiment sketched in Fig.~\ref{Fig_2} we take the particular case of a single--mode transmission line with a time--varying termination.  Due to its time translational symmetry, a static termination can be characterized in terms of a time domain reflection response $R$ that is a function solely of the time \emph{difference} between incidence ($t'$) and re--emission ($t$), i.e. $R(t-t')$.  By contrast, a time--varying termination has a response function that depends explicitly on when the wave is re-emitted, $\bar{R}(t,t-t')$.  As discussed in~\cite{Horsley2023}, we can alternatively define the exact same response as an explicit function of the \emph{incidence} time without changing any predictions.  For our time--varying termination we have the following relation between time domain input field $\bar{u}(t)$ and output field $\bar{v}(t)$,
\begin{align}
    \bar{v}(t)&=\int_{-\infty}^{t}{\rm d}t'\,\bar{R}(t,t-t')\,\bar{u}(t')\nonumber\\
    \to v(\omega)&=\int\frac{{\rm d}\omega'}{2\pi}R(\omega-\omega',\omega')\,u(\omega')\label{eq:tvm-input-output},
\end{align}
where we use a bar to denote a quantity in the time domain (omitting this for frequency domain quantities) and perform a Fourier transform into the frequency domain in the second line.  Assuming the modulation of the interface is periodic with period $T$, the angular frequencies ($\omega$) are spaced by $2\pi/T$ and the convolution in Eq.~(\ref{eq:tvm-input-output}) can be written as a matrix multiplication, with a reflection matrix defined analogous to Eq.~(\ref{eq:scattering-matrix}),  
\begin{equation}
    \boldsymbol{v}=\boldsymbol{R}\cdot\boldsymbol{u},\label{eq:reflection-matrix}
\end{equation}
as described theoretically in~\cite{Horsley2023}.  Note that, unlike the scattering matrix defined in (\ref{eq:scattering-matrix}), the above reflection matrix does not account for the waves transmitted through the termination and radiated into free space, which are here equivalent to a loss channel.  Note that in a static medium, the frequency domain reflection operator is proportional to a delta function, making $\boldsymbol{R}$ a diagonal matrix.  By contrast, in the time-varying medium investigated here, the reflection matrix is dense implying strong coupling between frequencies.  Once measured, $\boldsymbol{R}$ enables `waveform shaping' -- the specification of the input waveforms to optimise chosen features of temporally scattered output waves'.\\

\noindent{\bf Experimental setup and reflection matrix measurement}

\noindent Figure~\ref{Fig_2}(a) shows a schematic of our experiment. We form an ultrafast TVM at the end of our transmission line from a ring resonator incorporating two electronically modulated varactor diodes. Applying a reverse bias DC voltage across the diodes alters their capacitance, thereby changing the resonance frequency of the resonator. We inductively couple radiation to the resonator using the near-fields of a circular loop antenna, excited by input signals with frequencies ranging from \SI{1}{\mega\hertz} to \SI{1}{\giga\hertz}.  By design, the loop antenna is much smaller than the free-space wavelengths of these signals, thus radiating very poorly. As a consequence, energy is predominantly either absorbed within the resonator, or reflected back along the antenna connection, constituting the output signal. 

We first characterise the response of the system while the voltage across the varactor diodes, which we refer to as the {\it control voltage}, is held at a particular level -- i.e. without inducing any frequency coupling. Figure~\ref{Fig_2}(b) shows the magnitude and phase of the reflected signal as a function of frequency, for a range of applied control voltages. As the control voltage is tuned from 0-\SI{10}{V}, the resonance frequency of the circuit shifts approximately linearly between $\sim$0.2-\SI{0.8}{\giga\hertz}.

By rapidly dynamically modulating the control voltage, our system can be understood as a waveguide terminated by a time-varying complex impedance, where input signals are `scattered' into reflected output signals containing new frequency components.  In our experiments, we apply periodic control voltage modulations, with a period of \SI{1}{\micro\second}.  For a particular control voltage modulation, we measure the reflection matrix defined in Eq.~(\ref{eq:reflection-matrix}) of our TVM in a positive frequency Fourier basis by sequentially injecting continuous wave waveforms, of frequency ranging from \SI{1}{\mega\hertz} to \SI{1}{\giga\hertz} in \SI{1}{\mega\hertz} steps, recording the reflected waveform in the time domain, and subsequently Fourier transforming this signal to obtain the complex spectrum. In this way, the reflected spectrum corresponding to the $n^{\text{th}}$ input signal forms the $n^{\text{th}}$ column of the frequency-frequency reflection matrix, which as above we refer to as $\boldsymbol{R}$. See Methods for more details.

\section{Results}

\noindent{\bf Distortion-free eigenpulses}

\begin{figure*}[ht!]
    \includegraphics[width=0.98\textwidth]{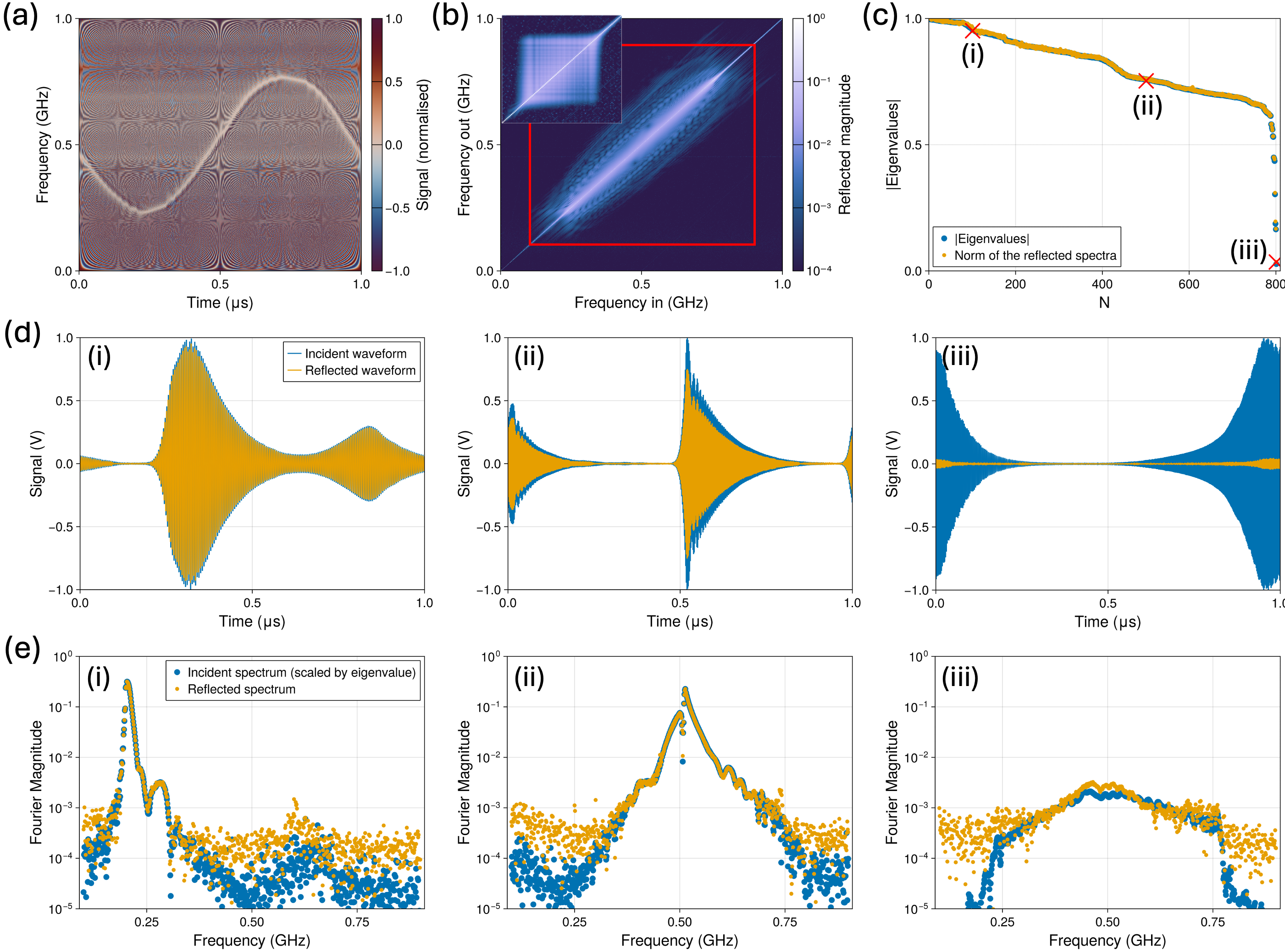}
    \caption{\textbf{Eigenpulses of a sinusoidally modulated time-varying medium.} \textbf{(a)} The real part of the frequency-to-time reflection matrix (downsampled), clearly showing the $1$\,MHz sinusoidal modulation of the control voltage. \textbf{(b)} Magnitude of the frequency-to-frequency reflection matrix calculated as the Fourier transform of (a).  Each pixel corresponds to a scattering of $f \pm n f_m$. For calculation of the eigenpulses only the area within the red bounded box is included. Inset: an equivalent reflection matrix for a $f_m=10$ MHz sinusoidal modulation. \textbf{(c)} The magnitude of the eigenvalues of the reflection matrix bounded by the red box in (b). Also shown are the norms of the reflected spectra for each eigenpulse. \textbf{(d)} Single periods of the waveforms of the injected and reflected eigenpulses corresponding to the 3 eigenvalues shown in red in (c). \textbf{(e)} The corresponding spectra after Fourier transforming the waveforms in (d). The incident spectra are scaled by the corresponding eigenvalue.}
    \label{Fig_3}
\end{figure*}

\noindent Once the reflection matrix of the TVM has been measured, its eigen-decomposition reveals a set of `eigenpulses' that are reflected as spectrally unchanged, but scaled by their corresponding eigenvalue $\lambda$, as recently described theoretically in~\cite{Horsley2023}.  More specifically, injecting the waveform $\boldsymbol{q}_n$ -- the $n^{\text{th}}$ eigenvector of $\boldsymbol{R}$ -- should result in a reflected waveform of the same form: ${\boldsymbol{v}_n= \boldsymbol{R}\cdot\boldsymbol{q}_n = \lambda_n\boldsymbol{q}_n}$.

To study such eigenpulses experimentally, we first set the control voltage to a sinusoidal modulation of \SI{1}{\mega\hertz}. Figure~\ref{Fig_3}(a) shows the time-domain measured signals that upon Fourier transforming results in our reflection matrix, \textbf{R} -- the magnitude of which is shown in Figure~\ref{Fig_3}(b). Since we have chosen a base frequency step of $f_{\text{min}}$=\SI{1}{\mega\hertz} for both the control voltage modulation, and the signal injection, and every other signal is a multiple of this, each off-diagonal element within $\boldsymbol{R}$ represents a scattered side-band. As in spatial diffraction, $1^{\text{st}}$ order scattering is strongest with subsequent higher orders becoming successively weaker. For our relatively low modulation frequency of \SI{1}{\mega\hertz}, this means that the significant values in the reflection matrix are found close to the main diagonal. Inset in Fig.~\ref{Fig_3}(b) we also show an example of the reflection matrix obtained for a \SI{10}{\mega\hertz} sinusoidal modulation. Due to the larger modulation frequency the side bands extend further from the diagonal.  

Figure~\ref{Fig_3}(c) shows the modulus of the eigenvalues calculated from the region of the reflection matrix highlighted within the red box in Fig.~\ref{Fig_3}(b).  Inclusion of the parts of the reflection matrix outside of this red box contributes only an additional set of eigenvectors that are near single frequency, since the matrix is very close to diagonal in these regions.  Meanwhile, as shown in the supplementary material, for a slow modulation we can write the pulse spectrum as $\exp({\rm i}S(\omega))$ so that, similar to semiclassical quantum mechanics where the local momentum is given as the gradient of the phase, the time coordinate becomes the frequency derivative of the phase, $t=\partial_{\omega}S$.  In this asymptotic limit an eigenpulse with eigenvalue $\lambda$ corresponds to a constant value of the reflection response function (\ref{eq:tvm-input-output}), $R(t,\omega)=R(\partial_{\omega}S,\omega)=\lambda$.  In the supplementary material we show that the solution of this equation for the function $S(\omega)$ gives rise to finite duration pulses temporally localized where the time modulation of the resonance frequency is linear, as indicated in the experimental results shown in Fig.~\ref{Fig_3}d.

To test whether the eigenpulses do indeed retain the same spectrum after interaction with our TVM, we inject the waveforms specified by three eigenvectors (highlighted on Fig.~\ref{Fig_3}(c)), and measure the corresponding reflected waveform in each case. These input and reflected waveforms are shown in the time domain in Figs.~\ref{Fig_3}(d)(i-iii), alongside the magnitude of their Fourier spectra which are shown in Figs.~\ref{Fig_3}(e)(i-iii). Here the injected waveform spectra have been scaled by the magnitude of the corresponding eigenvalue.

Figures~\ref{Fig_3}(d-e) show that when injecting waveforms corresponding to the eigenvectors of the reflection matrix, the reflected waveforms are very similar to those inputted. This is most clearly seen from the spectra in Fig.~\ref{Fig_3}(e) in which the reflected spectrum closely overlies the scaled injected spectrum for all magnitudes greater than $\sim10^{-3}$ (approximately our noise floor). The norm of the reflected spectrum vectors should also equal the eigenvalues, and this comparison is included in Fig.~\ref{Fig_3}(c), showing excellent agreement. A video showing the eigenpulses corresponding to all of the eigenvalues of the reflection matrix can be found in the supplementary material.

\begin{figure}[t!]
    \includegraphics[width=0.48\textwidth]{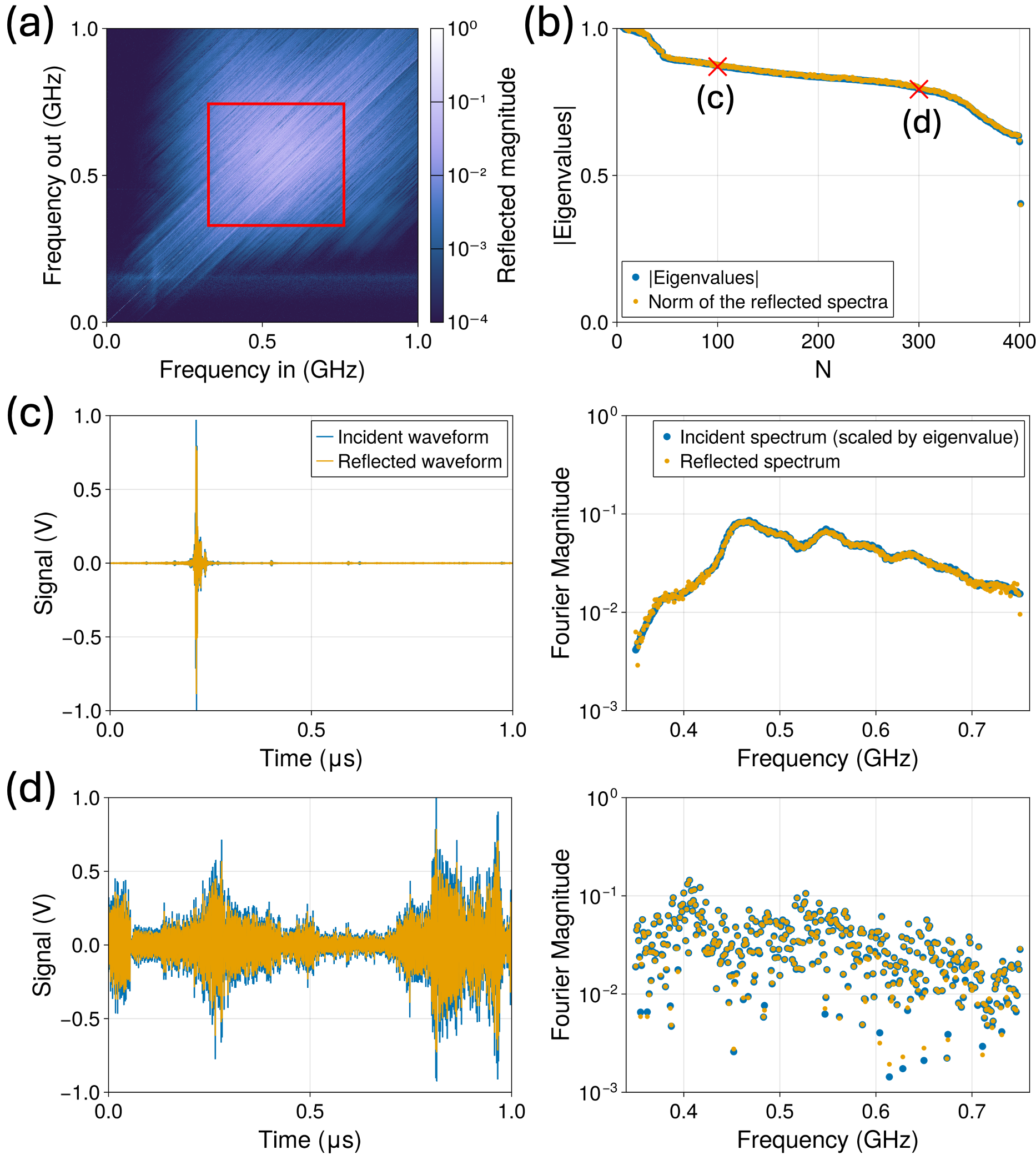}
    \caption{\textbf{Eigenpulses of a randomly modulated time-varying medium.} \textbf{(a)} The magnitude of the frequency-to-frequency reflection matrix for a `randomly' modulated time-varying medium. \textbf{(b)} The magnitude of the eigenvalues of the region of the reflection matrix bounded by the red box in (a). \textbf{(c)} Time domain waveforms and frequency spectrum of the injected and reflected eigenpulses for the larger magnitude eigenvalue shown in red in (b). \textbf{(d)} As for (c), but for the smaller magnitude eigenvalue shown in (b).}
    \label{Fig_4}
\end{figure}

Taking a more extreme case, we next investigate the excitation of eigenpulses within a `randomly' modulated TVM: here the control voltage signal is formed from 100 frequency components between 1 and \SI{100}{\mega\hertz} in \SI{1}{\mega\hertz} steps, each of which have a randomly chosen amplitude and phase (uniformly distributed between 0-\SI{10}{V}, and 0-2$\pi$\, rads., respectively). The eigenvalues and norms of the reflected spectra are shown in Fig.~\ref{Fig_4}(b) for the region of the reflection matrix shown in Fig.~\ref{Fig_4}(a) bounded by the red box. The reflection matrix within this region is dense, indicative of strong scattering between frequencies. Reflection from this modulated interface takes an arbitrary incident signal and scrambles it. Yet, as shown in Fig.~\ref{Fig_4}(c) and (d), when we select two of the eigenvalues and inject their corresponding eigenpulses, these are once again found to maintain their spectral content. This result shows that in our test system, eigenpulses can be created regardless of the modulation complexity of the time-varying medium. Due to the more complex modulation, in this case we generally find the eigenpulses to have a broader range of spectral components leading to a richer variety of pulse shapes. All eigenpulses of this random scattering system and their spectra can also be seen in the corresponding video in the supplementary material.\\

\begin{figure}[t!]
    \includegraphics[width=0.48\textwidth]{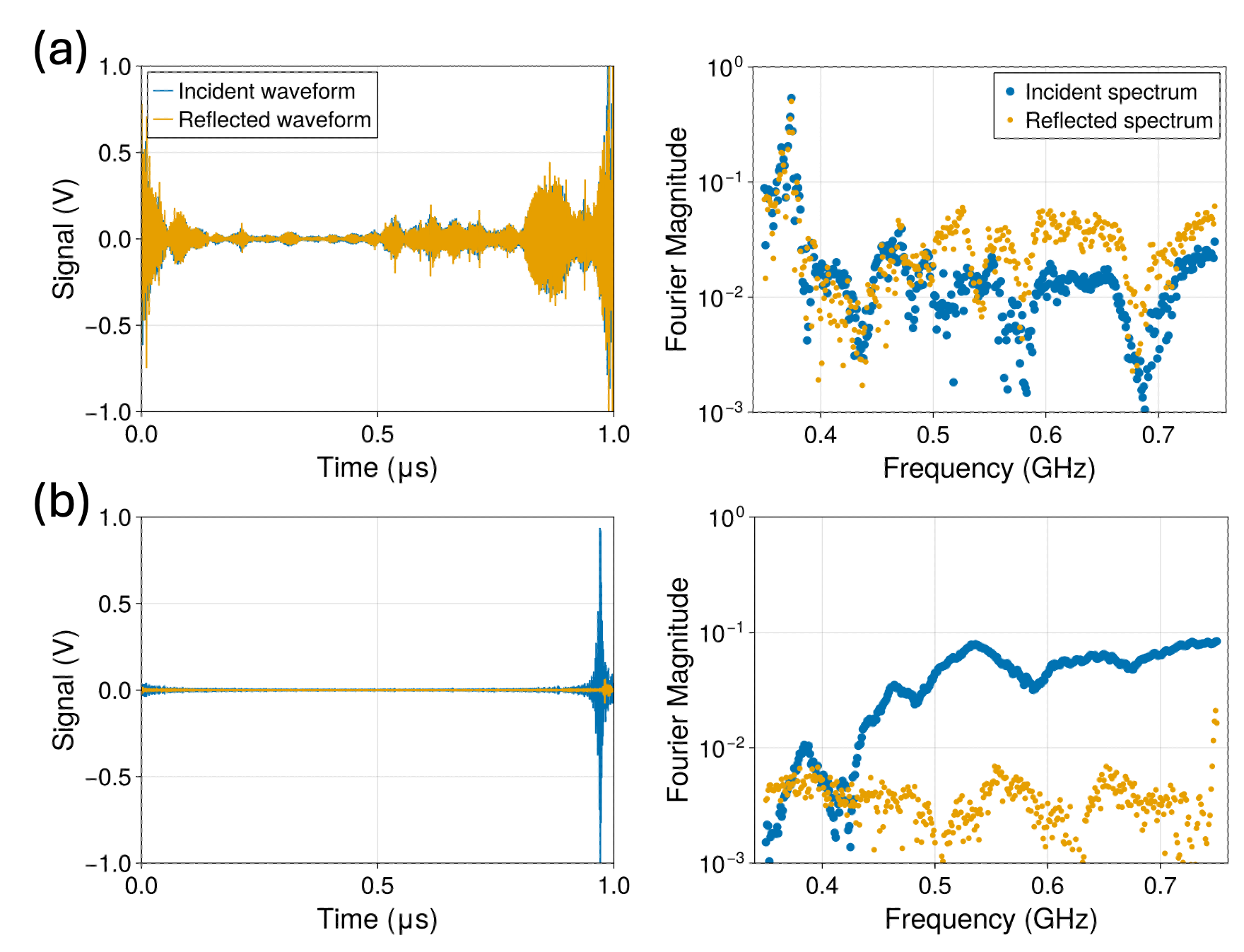}
    \caption{\textbf{Maximally and minimally reflected waveforms.} The waveforms and spectra of the maximally \textbf{(a)} and minimally \textbf{(b)} reflected waveforms, corresponding to the largest and smallest eigenvalues of $\boldsymbol{R}^\dagger\cdot\boldsymbol{R}$ for the reflection matrix shown in Fig.~\ref{Fig_3}(a).}
    \label{Fig_5}
\end{figure}

\noindent{\bf Minimally and maximally reflected pulses}

\noindent We now consider the design of input waveforms that maximise or minimise the total reflected power. While the eigenpulses described above display a range of reflected power levels (as shown in Figs.~\ref{Fig_3}(c) and~\ref{Fig_4}(b)), they are specifically designed to conserve the spectrum of the pulse, and so do not necessarily maximise or minimise power transfer. The total reflected power, $P$, can be written as
\begin{equation}\label{Eq_power}  
    P=\boldsymbol{v}^\dagger\cdot\boldsymbol{v}\,=\boldsymbol{u}^\dagger \cdot\boldsymbol{R}^\dagger\cdot\boldsymbol{R}\cdot\boldsymbol{u}.
\end{equation}
Equation~\ref{Eq_power} indicates that the injected waveform that will maximise the reflected power ($P_{\text{max}}$) is given by the eigenvector, $\boldsymbol{q}_{\text{max}}$, of matrix ${\boldsymbol{R}^\dagger\cdot\boldsymbol{R}}$ associated with the maximum eigenvalue $\lambda_{\text{max}}$ (noting that matrix ${\boldsymbol{R}^\dagger\cdot\boldsymbol{R}}$ is Hermitian with real eigenvalues). Thus ${P_{\text{max}} = \boldsymbol{q}_{\text{max}}^\dagger\cdot\boldsymbol{R}^\dagger\cdot\boldsymbol{R}\cdot\boldsymbol{q}_{\text{max}} = \lambda_{\text{max}}}$. Analogously, the waveform resulting in the minimum reflected power is given by the eigenvector of ${\boldsymbol{R}^\dagger\cdot\boldsymbol{R}}$ associated with the minimum eigenvalue. We note this approach is equivalent to calculation of the singular value decomposition of the reflection matrix, and injecting waveforms associated with the extreme singular values (see Methods).

The injected and reflected waveforms and spectra corresponding to the maximum and minimum eigenvalues of ${\boldsymbol{R}^\dagger\cdot\boldsymbol{R}}$, for the randomly modulated TVM, are shown in Fig.~\ref{Fig_5}. The smallest eigenvalue is 0.017 (the matrix is normalised such that the largest eigenvalue is unity) indicating that over 98\% of incident power is absorbed over a broad bandwidth. Videos showing waveforms and spectra for all of the eigenvalues of ${\boldsymbol{R}^\dagger\cdot\boldsymbol{R}}$ for both the sinusoidally and randomly modulated time-varying medium can be found in the supplementary material.\\

\noindent{\bf Optimal spectral focusing of reflected waveforms}

\begin{figure*}[ht]
    \includegraphics[width=0.98\textwidth]{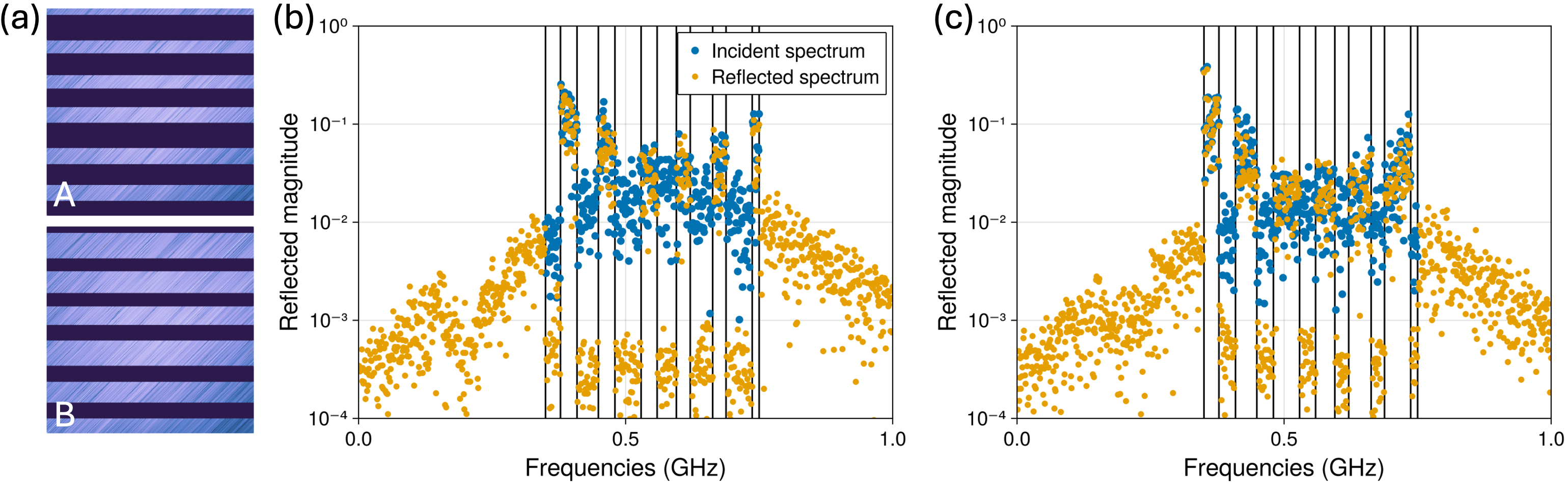}
    \caption{\textbf{Frequency domain `waveform shaping'.} (\textbf{a}) We form two masked reflection matrices, $\mathbf{R}_\textrm{A}$ and $\mathbf{R}_\textrm{B}$, from the reflection matrix bounded by the red box in Fig.~\ref{Fig_4}. The areas masked in $\mathbf{R}_\textrm{A}$ are unmasked in $\mathbf{R}_\textrm{B}$, and vice versa. \textbf{(b)} We find the largest eigenvalue of the contrast operator (Eqn.~\ref{eq:contrast}) and inject the waveform constructed from the corresponding eigenvector. The spectrum of the reflected waveform maximises the power within the unmasked frequency bands of $\mathbf{R}_\textrm{A}$, while maximising the contrast with the unmasked regions of $\mathbf{R}_\textrm{B}$. \textbf{(c)} As for (b), but maximising the reflected power in the unmasked frequency bands of $\mathbf{R}_\textrm{B}$. }
    \label{Fig_6}
\end{figure*}

\noindent Finally we demonstrate how to design input waveforms that are concentrated to specific user-defined frequency bands upon reflection from our TVM. This is analogous to techniques in spatial wavefront shaping, where optical power is focussed to specific spatial locations at the output of a static complex medium.  Here we follow the recent work of Shaughnessy et al.~\cite{Shaughnessy2024}, defining a `contrast operator' $\boldsymbol{\mathcal{C}}$, given by
\begin{equation}
    \boldsymbol{\mathcal{C}}=\left(\boldsymbol{R}^\dagger_B \cdot\boldsymbol{R}_B\right)^{-1}\cdot\boldsymbol{R}^\dagger_A\cdot \boldsymbol{R}_A.\label{eq:contrast}
\end{equation}
The `target' spectral regions we wish to focus energy to are specified through the partial reflection matrices $\boldsymbol{R}_A$ and $\boldsymbol{R}_B$. Here ${\boldsymbol{R}_{A}=\boldsymbol{\Pi}_{A}\cdot\boldsymbol{R}}$ is a masked version of the full reflection matrix, where matrix $\boldsymbol{\Pi}_{A}$ is a masking operator which leaves the target spectral output channels unchanged (region $A$), while setting all other output frequency channels to a reduced value, where we want the intensity to be suppressed (region $B$). The second matrix  $\boldsymbol{R}_{B}=\boldsymbol{\Pi}_{B}\cdot\boldsymbol{R}$ masks out the complementary spectral regions of the reflection matrix. See Methods for the construction of matrices $\boldsymbol{\Pi}_{A}$ and $\boldsymbol{\Pi}_{B}$. Note that in cases where the masks reduce any spectral region to zero, the inverse in (\ref{eq:contrast}) must be computed as a pseudo--inverse.

Despite first appearances, due to the positive definiteness of the operators ${\boldsymbol{R}_{A,B}^{\dagger}\cdot\boldsymbol{R}_{A,B}}$, the contrast operator defined in Eq.~(\ref{eq:contrast}) equals a Hermitian operator after the similarity transformation ${\boldsymbol{\mathcal{C}}'=(\boldsymbol{R}_{B}^{\dagger}\cdot\boldsymbol{R}_{B})^{1/2}\cdot\boldsymbol{\mathcal{C}}\cdot(\boldsymbol{R}_{B}^{\dagger}\cdot\boldsymbol{R}_{B})^{-1/2}}$.  Therefore, as discussed in~\cite{Shaughnessy2024}, the eigenvector of $\boldsymbol{\mathcal{C}}$ with the largest eigenvalue maximizes the contrast between the power in the spectral regions A and B.

To experimentally test this approach, we modify the reflection matrix measured for the randomly modulated TVM, specifying an arbitrarily chosen set of frequencies bands we wish to focus reflected power into. The resulting matrices $\boldsymbol{R}_A$ and $\boldsymbol{R}_B$ are shown in Fig.~\ref{Fig_6}(a). We then determine the eigenvector associated with the largest eigenvalue of the contrast operator $\boldsymbol{\mathcal{C}}$, and propagate this waveform through our TVM. The injected and reflected spectra are shown in Fig.~\ref{Fig_6}(b), whilst in Fig.~\ref{Fig_6}(c) we have performed the same process but reversed the matrices in the contrast operator.

The calculated input waveforms contain spectral components distributed throughout the frequency range of the reflection matrix. Upon propagation through our time-varying medium, we find reflected energy is concentrated into the target spectral bands, or scattered outside of the frequency range of the partial reflection matrices altogether. Indeed, by applying the contrast operator we have over 4 orders of magnitude more power in our target frequency bands compared to those which we wished to minimise.  We note that the magnitude of the spectral components in the region which we wished to minimise are at approximately the noise floor of our experiment.

\section{Summary and Conclusions}

We have experimentally measured the reflection matrix of a strongly frequency--scattering time-varying medium, extending the concept of `wavefront shaping' to `waveform shaping' in \emph{temporally} complex media.  Although an arbitrary input waveform is heavily distorted by both the dispersion and time modulation of the transmission line termination, we have unveiled a family of waveforms, termed eigenpulses, that -- upon reflection -- are spectrally unchanged, the time modulation of the termination acting to precisely counteract its dispersion.  Similarly, we have used the reflection matrix to calculate the waveforms that will be maximally and minimally reflected by our system, demonstrating strong absorption across a broad input bandwidth, despite the narrow linewidth of the unmodulated resonance.  

As a further analogue of `wavefront shaping', we implemented the temporal equivalent of focussing through a disordered medium, finding the waveform that will maximally `focus' power into specific spectral bands whilst minimising it in others.  We emphasize that, unlike an ordinary filter that achieves a similar effect through removing part of the signal, here we are \emph{converting} part of the spectrum into these desired frequency bands.  This difference enables us to derive a special set of pulses that are maximally amplified or converted into a given frequency band, a phenomenon that could be useful in applying TVM to concentrate energy into a given set of output modes.

Using non--linear optical modulation (see e.g.~\cite{tirole2022}), our scattering matrix characterization of a TVM could also be extended to optical frequencies.  Besides fundamental interest, we note that, once determined, the eigenvectors can be applied to select out spectrally broad portions of a signal that are absorbed or reflected.  Furthermore, encoding information in the eigenbasis of a highly scattering TVM, the output amplitudes are simply scaled, as if the waveform was interacting with a static medium.  Assuming we don't have access to the reflection matrix, encoding information in any other basis will require significant signal processing to extract the same information after interaction with the TVM.   

Our demonstrated application of the reflection matrix to describe a TVM opens up several possibilities.  The most direct is to apply the techniques of inverse design~\cite{molesky2018,garg2025}, determining the parameter (i.e., control voltage) modulation necessary to realise a desired set of eigenpulses that are e.g. perfectly absorbed or reflected, or transformed in a specified way, rather than choosing a modulation and then determining the eigenpulses as shown above.  Taking this further, we can think of a TVM as an unusual kind of filter, generalizing existing static photonic designs~\cite{small2012,miller2025} to include dynamic media, the filter potentially absorbing a range of temporal pulse shapes, leaving some desired set unchanged, or simply scaled.  In contrast to a standard filter that simply masks a region of the input spectrum, a TVM is sensitive to the phase and magnitude of the input spectral amplitudes.

Another implication of our results comes from understanding the eigenpulses of a dispersive TVM as being special fields where the material time modulation precisely counteracts the dispersion of the medium, leaving the pulse shape unchanged.  In a static medium the eigenpulses are simply continuous waves, every polychromatic pulse undergoing a reshaping due to the frequency dependence of the material parameters.  Dispersion compensation is an important topic in fibre optic communications~\cite{ouellette1987} and interferometry~\cite{resch2007}, and it is interesting that this occurs naturally for the eigenpulses of a TVM.

Finally we can also link our results for a randomly generated material modulation shown in Fig.~\ref{Fig_4} to the subject of chaotic communication~\cite{pecora1990,argyris2005,zaminga2025}, where two separated chaotic sources are synchronized with a common drive.  Just as this synchronization provides the `key' necessary to decode any signal imprinted on this chaotic light, here the knowledge of the eigenbasis of the reflection matrix could be used to encode/decode a signal incident onto a randomly generated material modulation.

\section{Methods}

\subsection{Experimental setup}

We form our time-varying medium from a ring resonator incorporating two varactor diodes (Infineon Technologies BB833E6327HTSA1). The ring is formed from \SI{35}{\um} copper on a \SI{1.5}{\mm} thick Rogers R4350B substrate using standard PCB processing methods. The outer diameter of the ring is \SI{27}{\mm} and the track width is \SI{1.3}{\mm}, with \SI{1.3}{\mm} diametrically opposed gaps in the ring across which the diodes are soldered -- see Fig.~\ref{Fig_2}(a). Applying a reverse bias voltage across the diodes alters their capacitance, thereby altering the resonance frequency of the resonator.

We inductively couple radiation to the resonator using the near-fields of a \SI{25}{\mm} diameter circular loop antenna formed from \SI{1}{\mm} diameter copper wire connected between the central core and shielding of a coaxial cable. By altering the distance between the ring resonator and the exciting antenna, the coupling strength between them can be altered. We use a ZX30-9-4-S+ directional-coupler from Mini Circuits connected to the coaxial cable to enable us to isolate and measure reflected waveforms using a Keysight DSO-Z-254A oscilloscope, with injected waveforms being generated by a Keysight M8195A arbitrary waveform generator. Since the resonator and loop antenna are much smaller than any free-space wavelengths used ($\sim\lambda/15$ at our highest frequency of \SI{1}{\giga\hertz}) it will radiate very poorly. As such any energy lost to reflection will be assumed to be due to absorption (predominantly within the diodes), and we can essentially consider our system as a waveguide terminated by a time-varying complex impedance.

\subsection{Control voltage modulation} 

The sinusoidal modulation of the control voltage (Fig.~\ref{Fig_3}) was at a frequency of \SI{1}{\mega\hertz}, while the `random' modulation (Fig.~\ref{Fig_4}) was formed from 100 frequency components between 1 and \SI{100}{\mega\hertz} in \SI{1}{\mega\hertz} steps, each of which have a random amplitude and phase. As such, both modulations are periodic with a period of \SI{1}{\micro\second}. All modulations used here have a \SI{5}{\volt} DC offset with a peak-to-peak modulation of \SI{10}{\volt}.

\subsection{Reflection matrix measurement}

To generate our reflection matrices we begin by injecting consecutive sinusoidal and cosinusoidal waveforms from \SI{1}{\mega\hertz} to \SI{1}{\giga\hertz} in \SI{1}{\mega\hertz} steps and measuring the reflected waveforms, before combining the two data sets (${\boldsymbol{R}_{t,f}=\boldsymbol{R}_{t,f}^{\rm (cos)}+{\rm i}\boldsymbol{R}^{\rm (sin)}_{t,f}}$).  To remove low frequency signals arising from our modulating bias voltage (the modulation also acts as a secondary source) we subsequently inject the same waveforms but $\pi$ out of phase, and subtract the two.

The measured reflection matrix $\boldsymbol{R}_{t,f}$ thus has inputs $\boldsymbol{u}$ in the positive frequency Fourier basis, and output channels that are quadrature time-series measurements of our reflected waveforms. A downsampled example for a \SI{1}{\mega\hertz} sinusoidal modulation of the bias voltage is shown in Fig.~\ref{Fig_3}(a), where the expected variation in resonance frequency as a function of time is clearly evident, i.e. the reflection null tracks the sinusoidal modulation.

Frequency-to-frequency reflection matrices, $\boldsymbol{R}$, defined in Eq. (\ref{eq:reflection-matrix}), are obtained by Fourier transforming the output time axis in the measured $\boldsymbol{R}_{t,f}$ reflection matrix.  Since we have chosen a base frequency of \SI{1}{\mega\hertz}, every other frequency (whether injected or modulating) is a multiple of this, and each pixel within our $\boldsymbol{R}$ matrices represents a scattered side-band.  For the case of a sinusoidal modulation, the $\boldsymbol{R}$ matrix calculated from our measured reflection data is shown in Fig.~\ref{Fig_3}(b).

For the spectral focusing experiments we choose a different distance between the loop antenna and the ring resonator. Here, rather than wanting to maximise the absorption within the resonator, we wish to maximise the scattering between frequencies. To do this we want to maximise the coupling of radiation to the resonator, which we achieve by minimising the gap between them. This moves us into the over--coupled regime where the resonance takes the form of a broader shallower resonance with a strong phase variation as a function of frequency. For the equivalent plots to Fig.~\ref{Fig_2}(b) for the over-coupled system see \S 3 in the supplementary material. The corresponding measured $\boldsymbol{R}_{t,f}$ and modulating voltage profile can be found in \S 4 of the supplementary material.

\subsection{Eigenpulse preparation}
Once the reflection matrix of the time-varying medium has been measured, its eigen-decomposition reveals a set of `eigenpulses' that are reflected spectrally unchanged, but scaled by their corresponding eigenvalue, as recently described theoretically in~\cite{Horsley2023}. More specifically, $\boldsymbol{R}$ may be factorised into the product of three matrices, ${\boldsymbol{R} = \boldsymbol{Q}\cdot\boldsymbol{\Lambda}\cdot\boldsymbol{Q}^\dagger}$, where the $n^{\text{th}}$ column of unitary matrix $\boldsymbol{Q}$ holds the $n^{\text{th}}$ eigenvector $\boldsymbol{q}_n$ (expressed as complex amplitudes in the Fourier basis), and $\boldsymbol{\Lambda}$ is a diagonal matrix holding the $n^{\text{th}}$ eigenvalue $\lambda_n$ in element $\Lambda_{n,n}$. Thus injecting the waveform formed from the $\boldsymbol{q}_n$ amplitudes results in a reflected waveform of ${\boldsymbol{v}= \boldsymbol{R}\cdot\boldsymbol{q}_n = \lambda_n\boldsymbol{q}_n}$.

\subsection{Maximising/minimising total reflected power}

The total reflected power, $P$, from our system can be written as
\begin{equation}\label{Eq_power2}  
    P=\boldsymbol{v}^\dagger\cdot\boldsymbol{v}\,=\boldsymbol{u}^\dagger \cdot\boldsymbol{R}^\dagger\cdot\boldsymbol{R}\cdot\boldsymbol{u}.
\end{equation}
Equation~\ref{Eq_power2} indicates that the injected waveform that will maximise the reflected power from our system ($P_{\text{max}}$) is given by the eigenvector, $\boldsymbol{q}_{\text{max}}$, of matrix ${\boldsymbol{R}^\dagger\cdot\boldsymbol{R}}$ associated with the maximum eigenvalue $\lambda_{\text{max}}$ (noting that matrix ${\boldsymbol{R}^\dagger\cdot\boldsymbol{R}}$ is Hermition with real eigenvalues). Thus ${P_{\text{max}} = \boldsymbol{q}_{\text{max}}^\dagger\cdot\boldsymbol{R}^\dagger\cdot\boldsymbol{R}\cdot\boldsymbol{q}_{\text{max}} = \lambda_{\text{max}}}$.

Equivalently, this input waveform can also be found by performing a singular value decomposition of the reflection matrix. In this case $\boldsymbol{R}$ may be factorised into the product of three matrices, ${\boldsymbol{R} = \boldsymbol{A}\cdot\boldsymbol{\Sigma}\cdot\boldsymbol{B}^\dagger}$, where $\boldsymbol{A}$ and $\boldsymbol{B}$ are unitary matrices, and $\boldsymbol{\Sigma}$ is a real-valued diagonal matrix holding the 
 singular values along its diagonal, such that $\Sigma_{n,n} = \sigma_n$. The $n^{\text{th}}$ column of $\boldsymbol{A}$ holds the $n^{\text{th}}$ left singular vector, $\boldsymbol{a}_n$, and the $n^{\text{th}}$ column of $\boldsymbol{B}$ holds the $n^{\text{th}}$ right singular vector, $\boldsymbol{b}_n$.
Injecting the waveform formed from the complex amplitudes contained within the right singular vector $\boldsymbol{b}_{\text{max}}$, associated with the maximum singular value $\sigma_{\text{max}}$, results in a reflected waveform of ${\boldsymbol{v} = \boldsymbol{R}\cdot\boldsymbol{b}_{\text{max}} = \sigma_{\text{max}}\boldsymbol{a}_{\text{max}}}$, and the reflected power is given by ${\boldsymbol{v}^{\dagger}\cdot\boldsymbol{v} = \sigma_{\text{max}}^2\boldsymbol{a}_{\text{max}}^\dagger\cdot\boldsymbol{a}_{\text{max}} = \sigma_{\text{max}}^2}$. Thus we see that ${\lambda_{\text{max}} = \sigma_{\text{max}}^2}$ and $\boldsymbol{q}_{\text{max}} = \boldsymbol{b}_{\text{max}}$.

\subsection{Construction of matrices $\boldsymbol{\Pi}_{A}$ and $\boldsymbol{\Pi}_{B}$}

To construct $\boldsymbol{\Pi}_{A}$, we start with an identity matrix $\boldsymbol{I}$ of dimension equivalent to the number of output spectral channels of the measured reflection matrix $\boldsymbol{R}$. We now replace with zero any diagonal elements of $\boldsymbol{I}$ that correspond to rows of $\boldsymbol{R}$ that we wish $\boldsymbol{\Pi}_{A}$ to set to zero. The resulting matrix is $\boldsymbol{\Pi}_{A}$, and 
$\boldsymbol{\Pi}_{B}$ is then given by ${\boldsymbol{\Pi}_{B}=\boldsymbol{I}-\boldsymbol{\Pi}_{A}}$.\\


\section{Acknowledgements}

I.R.H. and S.A.R.H. acknowledges support from the EPSRC via the META4D Programme Grant (EP/Y015673/1), and would like to thank Gregory Chaplain and Timothy Starkey for useful discussions.  D.B.P. and S.A.R.H. acknowledge additional support through the EPSRC grant ``Photon management in dynamic complex scattering media'' (EP/Z535928/1). D.B.P. also acknowledges financial support from the European Research Council (Consolidator Grant ``ModeMixer" 101170907).

\bibliographystyle{apsrev4-2}
\bibliography{bib}

\end{document}